\documentclass[twocolumn, aps, prmaterials, reprint]{revtex4-2}
\usepackage{amsmath,amssymb}
\usepackage{graphicx}
\usepackage{dcolumn}
\usepackage{multirow}
\usepackage{bm}
\usepackage[english]{babel}
\usepackage{adjustbox}
\usepackage{booktabs}
\usepackage{tabularx}
\usepackage{xcolor}
\definecolor{BLUE}{named}{blue}
\definecolor{BLUE}{named}{blue}
\usepackage{svg}
\usepackage{array}

\newcolumntype{R}[1]{>{\raggedleft\let\newline\\\arraybackslash\hspace{0pt}}m{#1}}

\usepackage{hyperref} 
\usepackage{cleveref} 
\begin{document}


\title{Coupling between small polarons and ferroelectricity in BaTiO\texorpdfstring{$_3$}{3}}

\author{Darin Joseph}
\affiliation{Dipartimento di Fisica e Astronomia, Universit\`{a} di Bologna, \texorpdfstring{$40127$}{40127} Bologna, Italy}

\author{Cesare Franchini}
\affiliation{University of Vienna, Faculty of Physics, Center for Computational Materials Science, Vienna, Austria}
\affiliation{Dipartimento di Fisica e Astronomia, Universit\`{a} di Bologna, 40127 Bologna, Italy}

\date{\today}

\begin{abstract}

In this study, we investigate the formation of electron and hole small polarons in the prototypical ferroelectric material BaTiO$_3$, with a focus on their interaction with ferroelectric distortive fields. To accurately describe the ferroelectric phase in electronically correlated BaTiO$_3$, we employ the HSE06 hybrid density functional, which addresses the limitations of conventional density functional theory (DFT) and Hubbard-corrected DFT+U models, providing a more precise depiction of both ferroelectric and polaronic behaviors. Our analysis spans three structural phases of BaTiO$_3$: cubic, tetragonal, and rhombohedral.
We uncover a unique phase-dependent trend in electron polaron stability, which progressively increases across the structural phases, peaking in the rhombohedral phase due to the constructive coupling between the polaron and ferroelectric phonon fields. In contrast, hole polarons exhibit a stability pattern largely unaffected by the phase transitions.
Furthermore, we observe that polaron self-trapping significantly alters the local ferroelectric distortive pattern, which propagates to neighboring sites but has a minimal effect on the long-range macroscopic spontaneous polarization. Charge trapping is also associated with localized spin formation, opening new possibilities for enhanced functionalities in multiferroic materials.
\end{abstract}

\maketitle

\section{Introduction}
\begin{figure*}[hbt]
\includegraphics[width=7in]{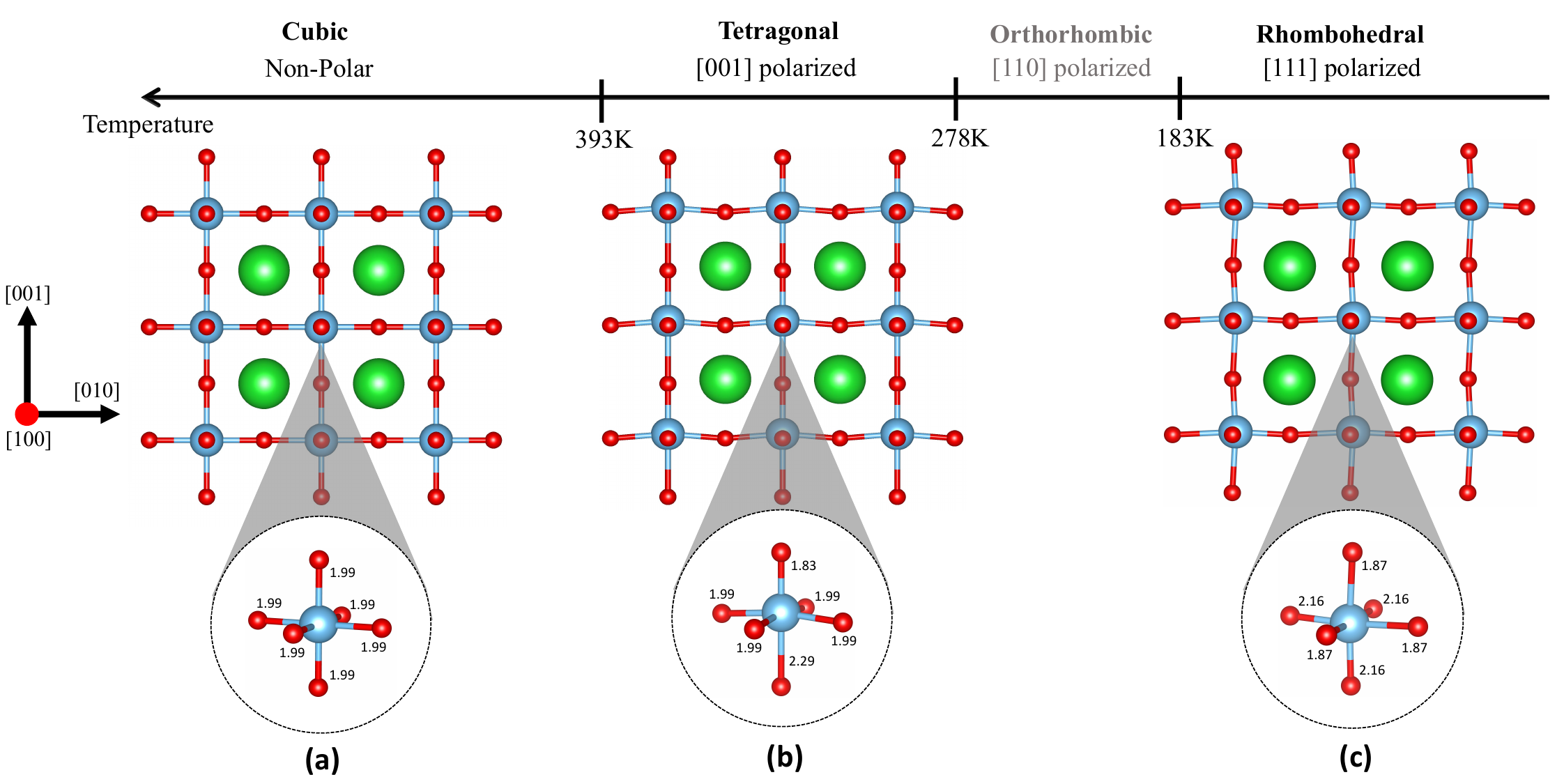}
\caption{(a) Cubic BaTiO$_3$ with zero Ti off-centering ($>$393K) (b) Tetragonal BaTiO$_3$ with Ti off-centering in [001] direction (393K - 278K) and (c) Rhombohedral BaTiO$_3$ with Ti off-centering in the [111] direction($<$183K). The orthorhombic phase is in the temperature range of 183-278K but its structure is not shown here because it was not considered in the present study. Detailed investigations of polaronic effects in the orthorhombic phase have been reported\cite{xu2019electron}. The insets represent the Ti-O bond-lengths in each phase.}
\label{fig:Polymorphs}
\end{figure*}

Marked by a spontaneous electric polarization (P$_s$) that can be reversed by an external electric field, ferroelectric (FE) materials are vital in a wide range of technological applications~\cite{cochran1960crystal,cohen1992origin,samara2001ferroelectricity,king1993theory}. Capitalizing on this unique switchable polarization property, their applications range from capacitors and transducers to nonvolatile memory devices~\cite{lines2001principles,scott2007applications,rabe2007physics}. Barium titanate (BaTiO$_3$) has been extensively studied, notably for its rich temperature-dependent phase transitions from the high-temperature cubic phase to the tetragonal, orthorhombic, and rhombohedral phases at lower temperatures~\cite{kwei1993structures,piskunov2004bulk}. This polymorphism opens avenues for investigating the correlation between the electronic structure and lattice distortions, as well as their evolution across different crystal symmetries.

\vspace{3mm}

The FE behavior and dielectric properties are altered due to the localized electronic states introduced by the presence of defects, such as oxygen vacancies or cation substitutions. In BaTiO$_3$ and related perovskites such as Pb(Zr${_x}$Ti$_{1-x}$)O$_3$, oxygen vacancies act as electron donors, potentially creating defect dipoles that pin domain walls, thereby affecting polarization switching~\cite{chen1994compositional,lou2009polarization,qiu2023extended,wang2016effects}. Similarly, for BaTiO$_3$ and SrTiO$_3$, titanium vacancies or antisite defects were shown to reduce or modify the ferroelectricity~\cite{michel2021interplay,das2014magnetic,klyukin2017effect,lee2015emergence,choi2009role}. Such defect-induced localization plays a pivotal role in tuning the electronic and ferroic properties of materials, thereby affecting both the static polarization and dynamic processes, such as charge transport. 

\vspace{3mm}

Similarly, polaron formation results in a localized state within the band structure. Unlike defects, which are associated with ionic or atomic anomalies, polarons are quasiparticles that form as a result of the interaction between charge carriers and lattice vibrations (phonons)~\cite{alexandrov2010advances,Landau:1933iwn,emin2013polarons,franchini2021polarons}. Polarons are usually classified into large and small polarons based on the extent of electron-phonon coupling, with small polarons showing strong coupling and localized behavior~\cite{alexandrov2010advances,Landau:1933iwn,emin2013polarons,franchini2021polarons,franchini2021polarons,sio2019polarons,holstein1959studies}. Polaron signatures have been detected in several materials, such as transition metal oxides, organic semiconductors, polymers, manganites, hybrid perovskites, cuprites, magnetic semiconductors, and 2D materials~\cite{reticcioli2020small,nagels1963electrical,crevecoeur1970electrical,stoneham2007trapping,zhugayevych2015theoretical,coropceanu2007charge,Birschitzky2025,roth2019foundations,hinrichs2018ellipsometry,de2016tracking,kaminski2002polaron,teresa1997evidence,daoud2002zener,zhou2000zener,cortecchia2017polaron,kang2018holstein}. Charge transport, colossal magnetoresistance, photoemission, surface reactivity, thermoelectricity, and (multi)ferroism are a few physical phenomena in which polaron-mediated effects play a crucial role~\cite{coropceanu2007charge,nelson2009modeling,reticcioli2017polaron,reticcioli2019interplay,teresa1997evidence,millis1996fermi,verdi2017origin,miyata2017large,miyata2018ferroelectric,Birschitzky2024}..

\vspace{3mm}

Theoretical and experimental studies have provided ample evidence for the presence of polarons in BaTiO$_3$. Experimental indications linked to the optical and transport properties of materials have long suggested the presence and impact of small polarons in BaTiO$_3$. For instance, in BaTiO$_3$ single crystals, the characteristic green luminescence has been linked to the presence of small polarons, and the transport behavior has been explained through hopping mechanisms involving these quasiparticles~\cite{aguilar1979x,ihrig1981conductivity,boyeaux1979small,iguchi1991polaronic,ihrig1978electrical,jing2017tuning}. Computational investigations using hybrid functionals have shown that hole polarons can self-trap on oxygen atoms in BaTiO$_3$ in both the tetragonal and cubic phases, consistent with the experimentally observed low-temperature photoluminescence~\cite{traiwattanapong2018self}. DFT+U studies also confirmed the stability of hole polarons in BaTiO$_3$’s cubic phase~\cite{erhart2014efficacy}. Understanding the interplay between the FE order and small polarons provides a unique opportunity to explore novel functionalities of this material. Gaining insights into whether and how FE distortions ($\delta_{FE}$) in BaTiO$_3$, affect polaron formation is essential for understanding the stability and behavior of polarons. In BaTiO$_3$, these distortions correspond to an off-centering of the Ti atom from the center of the TiO$_6$ octahedron, along with a small movement of the O atom, leading to a P$_s$. FE polarons result from such a constructive interplay and are described as polarons stabilized by $\delta_{FE}$. Such polarons have been shown to enhance the charge transport efficiency and optical performance of related materials, such as halide perovskites~\cite{miyata2017large,miyata2018ferroelectric}. Moreover, ongoing studies are examining multiferroism induced by small polarons in BaTiO$_3$ and similar perovskites, highlighting the broader significance of polaron-ferroelectric coupling in these oxides~\cite{tsunoda2019stabilization,xu2019electron,xu2024emergent}.

\vspace{3mm}

Despite the available results and knowledge, the interplay between ferroelectricity and small polarons in BaTiO$_3$ remains a topic that requires a deeper analysis. There is a limited number of literature and a detailed investigation needs to be done to get a clearer picture of the mutual influence of these phenomena. In particular, the impact of ferroelectricity on hole polarons in different phases of BaTiO$_3$ remains an unexplored domain. The inaccuracy of DFT+U in capturing the $\delta_{FE}$ complicates the theoretical investigation, since this method is used mostly to study polarons in materials~\cite{din2020electron,gebreyesus2023understanding}.

\vspace{3mm}

The main aim of this study is to fill this gap by investigating the formation of electron and hole polarons in three different phases of BaTiO$_3$ using hybrid functionals. The main focus of the current study is the analysis of the variation in polaron stability across different structures and the reciprocal impact of polarons on $\delta_{FE}$. By elucidating this complex interplay between ferroelectricity and small polarons across different phases and charge carrier types, our research advances the understanding of the complex interrelations between the FE order and charge localization in BaTiO$_3$. 

\section{Material background and computational approach}

As mentioned earlier, BaTiO$_3$ is a classic ferroelectric material that exhibits multiple structural phases with distinct polarization behaviors, as illustrated in Figure~\ref{fig:Polymorphs}. The cubic phase of BaTiO$_3$ has perfectly centered Ti and O atoms, possessing zero off-centering from the TiO$_6$ octahedron, while the ferroelectric tetragonal and rhombohedral phases are characterized by Ti off-centering along the [001] and [111] directions, respectively, leading to $\delta_{FE}$ in the systems. For the tetragonal phase, we calculated $\delta_{FE}$ to be $\sim0.16$~\AA\ along the [001] direction, while for the rhombohedral phase, $\delta_{FE}$ was computed to be $\sim0.10$~\AA\ along the [111] direction. 

\vspace{3mm}

The bond lengths vary significantly among the phases according to these distortions. In the cubic phase, all the bonds are equal in length. As shown in Figure~\ref{fig:Polymorphs}(b), the in-plane bonds of tetragonal BaTiO$_3$ are equal in length, whereas the out-of-plane bonds are of two different lengths, one short and the other long. The rhombohedral phase of BaTiO$_3$ also has two distinct bond types according to their length, three of which are longer and the other three are shorter, as shown in Figure~\ref{fig:Polymorphs}(c). The variation in bond lengths corresponds to the changes in the orbital hybridization between the Ti-3$d$ and O-2$p$ states, which plays a key role in stabilizing the ferroelectric phase in BaTiO$_3$ through enhanced covalent interactions. The modulation of this hybridization is fundamental to understanding the behavior of polarons and the FE stability of the material. 

\vspace{3mm}

While the orthorhombic phase of BaTiO$_3$ also exhibits ferroelectricity, the present study excludes this because its polaron behavior has been extensively studied previously\cite{xu2019electron}. Moreover, our focus on the tetragonal and rhombohedral phases is also due to their representation of uniaxial and triaxial polarization symmetries, respectively, which are more relevant to our study, which focuses on polarization-directional effects.

\vspace{3mm}

\begin{figure}
\includegraphics[width=\linewidth]{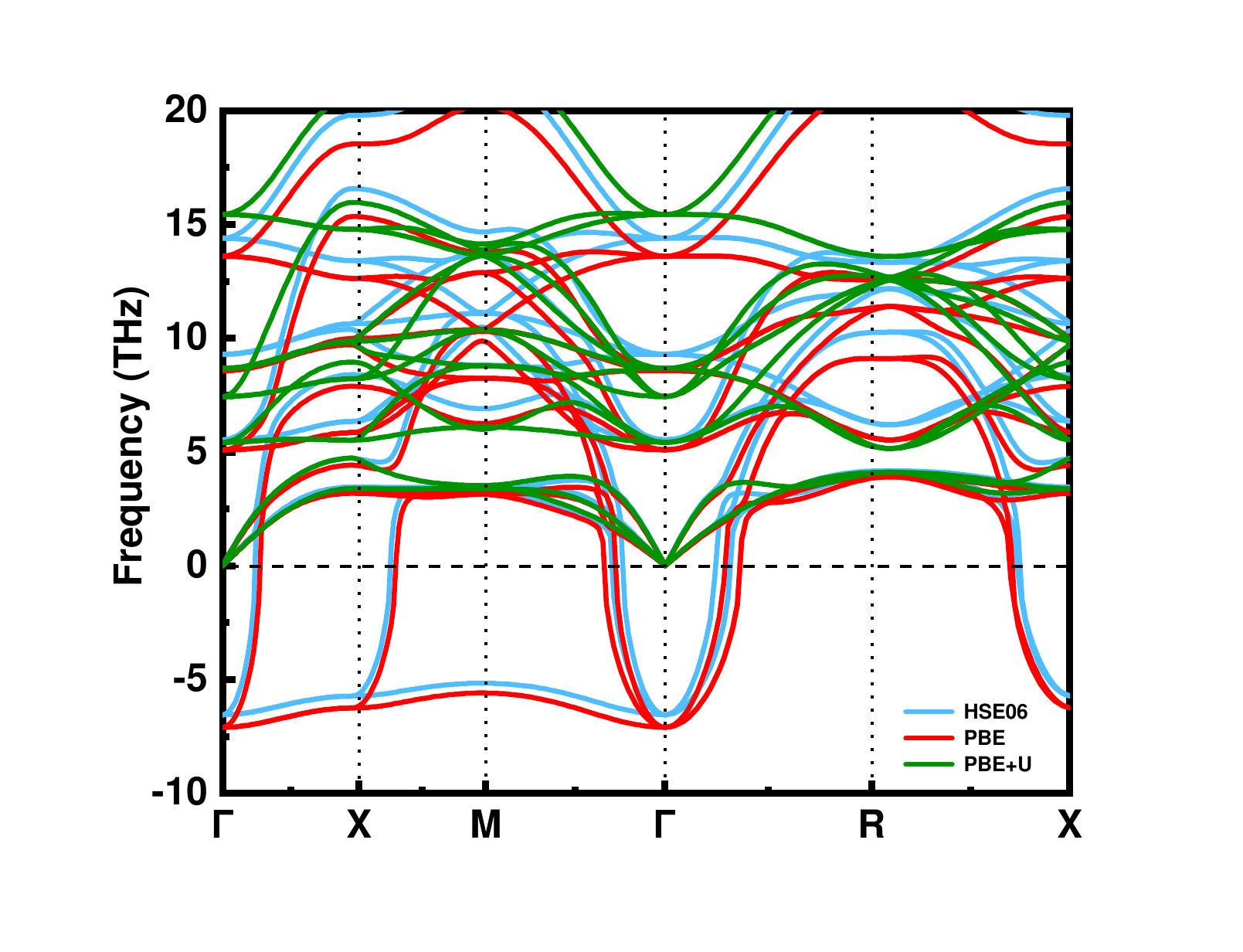}
\caption{Phonon dispersion curve of cubic BaTiO$_3$ plotted with HSE06, PBE and PBE+U. With PBE+U, for a U value of 6 eV, a clear vanishing of imaginary modes are observed. This depicts the effect of Hubbard U parameter in suppressing the $\delta_{FE}$ in the system.}
\label{fig:Phonon-dispersion curve comparison}
\end{figure}

From a computational perspective, Density Functional Theory (DFT) is a powerful tool for electronic structure calculations~\cite{hohenberg1964inhomogeneous,gross2013density}; however, the limitations of DFT are not unfamiliar, especially when it comes to the study of polarons~\cite{gavartin2003modeling,sadigh2014variational}. For modelling systems where electron localization is pivotal, DFT often leads to inaccuracies due to inherent self-interaction errors. The widely adopted approach to address this issue is the integration of the Hubbard U term into DFT calculations, which corrects these errors by accounting for the onsite coloumbic interaction~\cite{anisimov1991band,dudarev1998electron}. However, DFT+U poses challenges to the current study. The stabilization of $\delta_{FE}$ within FE systems is vital to this study. However, a linear reduction in $\delta_{FE}$ was observed for the FE phases of BaTiO$_3$ as the U value increased from 0 to 5 eV and beyond, resulting in a paraelectric cubic structure~\cite{din2020electron,gebreyesus2023understanding}. This observation was evident for both the tetragonal and rhombohedral phases during geometrical optimization. The stronger localization of electrons in the d-orbitals of the Ti atoms arises with the increase in the U value, consequentially diminishing the covalency due to a suppressed hybridization between the O-2$p$ and Ti-3$d$ orbitals. Figure~\ref{fig:Phonon-dispersion curve comparison} illustrates the phonon dispersion curve of cubic BaTiO$_3$ and clearly demonstrates that the PBE+U method does not show any imaginary phonon modes. The absence of these modes indicates that cubic BaTiO$_3$ does not exhibit soft phonon modes associated with FE instability. The inability of DFT+U to stabilize $\delta_{FE}$ has already been reported and studied\cite{din2020electron,gebreyesus2023understanding}. An alternative approach to overcome this limitation is to utilize hybrid functionals, thereby providing a balanced treatment to study polaronic phenomena while preserving the ferroelectricity in BaTiO$_3$’s FE phases~\cite{krukau2006influence}. 
\vspace{3mm}

Building on this understanding of the structural characteristics of BaTiO$_3$ and the validated computational approach, we detail the specific methodologies employed in our study to investigate polaron behavior within these ferroelectric phases.

\section{Methods}

We adopted the Heyd-Scuseria-Ernzerhof (HSE06) formalism~\cite{heyd2003hybrid}. The fraction of exact Hartree-Fock exchange was set to the standard value of $\alpha$ = 0.25. The first-principle calculations were carried out using  the Vienna Ab initio Simulation Package (VASP) in combination with the Projector Augmented Wave (PAW) approach~\cite{kresse1996efficient,perdew1996generalized,kresse1999ultrasoft,blochl1994projector}.The pseudopotentials used in our calculations are Ba$_{sv}$ (5s$^{2}$5p$^{6}$6s$^{2}$), Ti (3d$^{3}$4s$^{1}$), and O$_s$(2s$^{2}$2p$^{4}$). A plane wave energy cut-off of 320 eV was utilized in all calculations, with 10$^{-5}$ eV set as the convergence threshold of the electronic self-consistency. 
The polaronic systems were modelled with a 3x3x3 supercell, containing 135 atoms. 
The structural relaxation of the ionic positions was continued until the Hellman-Feyman forces were lower than 0.01 eV/~\AA, and a 2x2x2 Monkhorst-Pack k-point mesh was utilized for the brillouin zone integration. Employing HSE06 for lattice relaxations stabilized the $\delta_{FE}$ within the ferroelectric systems and concurrently provided lattice parameters and band gaps that more closely agreed to the experimental values\cite{kay1949xcv,kwei1993structures,shirane1952transition}. 

\vspace{3mm}

For the formation of electron polarons, an extra electron was introduced into the system and conversely for formation of hole polarons, an electron was removed. Simultaneously, a compensating background charge of opposite sign is included to keep the system neutral. This method can mimic the experimental conditions in which polarons are formed through photoexcitation rather than explicit defects, providing a computationally efficient and physically relevant approach for studying polaron formation and behavior\cite{carneiro2017excitation, jin2022photoinduced, molesky2018photoexcited, shelton2021thermally}. Initially, the result was a delocalised solution but by manually breaking the symmetry, it was possible to localize the electron and hole on the preferred site. For electron polarons, six Ti-O bonds around the desired Ti atom were extended, while for hole polarons, two Ti-O bonds next to the chosen O atom were extended. The polaronic formation energy was calculated using the following formula:

\begin{equation}
\begin{split}
E_{POL} = E_{dist}^{loc} - E_{unif}^{deloc}
\end{split}
\end{equation}

where (E$_{dist}^{loc}$) is the total energy of the supercell with a polaron, (E$_{unif}^{deloc}$) is the total energy of the pristine supercell with the extra charge delocalized in the entire crystal (no polaron). This formula allows us to calculate the energy gain of forming a polaron in the material as compared to a delocalized uniform solution~\cite{franchini2021polarons}. To further understand the energetics of polaron formation, we calculated the structural energy cost (E$_{ST}$) and electronic gain energy (E$_{EL}$). This helps decouple the contributions to the energy from lattice distortions and charge localization. They are calculated using the following formulas:

\begin{equation}
\begin{split}
E_{ST} = E_{dist}^{deloc} - E_{unif}^{deloc}
\end{split}
\end{equation}
\vspace{-10pt}
\begin{equation}
\begin{split}
E_{EL} = E_{dist}^{loc} - E_{dist}^{deloc}
\end{split}
\end{equation}

where (E$_{dist}^{deloc}$) is the system's total energy with delocalized charge contained in the lattice structure of the polaronic state\cite{reticcioli2020small}.

\begin{figure*}[hbt!]
\includegraphics[width=0.9\textwidth]{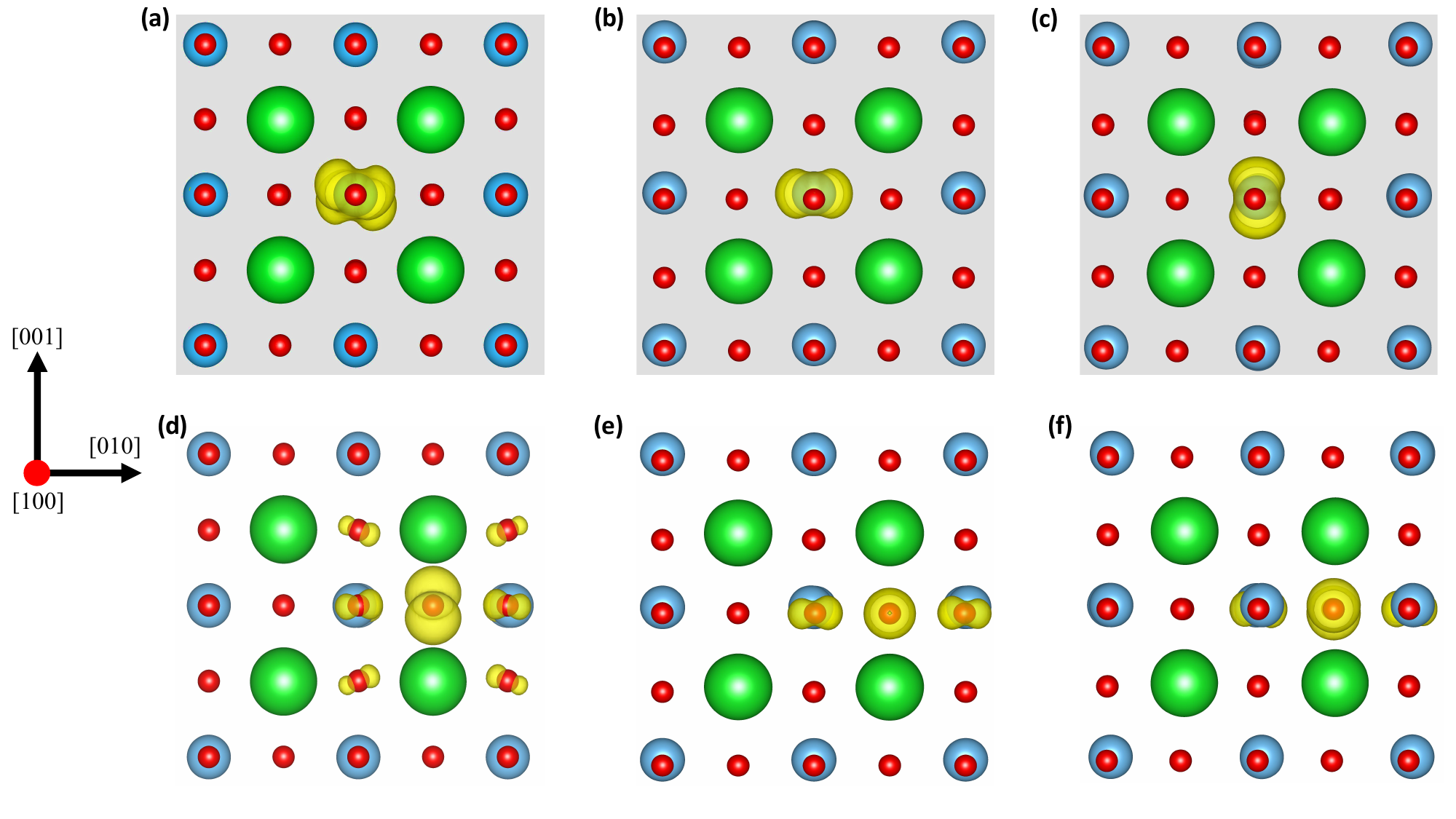}
\caption{The charge density plots presented in the top panel illustrate electron polaron isosurfaces in (a) Cubic BaTiO$_3$, (b) Tetragonal BaTiO$_3$, and (c) Rhombohedral BaTiO$_3$. Meanwhile, the bottom panel depicts hole polaron isosurfaces in (d) Cubic BaTiO$_3$, (e) Tetragonal BaTiO$_3$, and (f) Rhombohedral BaTiO$_3$.}
\label{Electron-polaron}
\end{figure*}
\vspace{3mm}

Berry-phase approach within the modern theory of polarization was employed to calculate the ferroelectric polarization of both the pristine and polaronic ferroelectric systems~\cite{resta2007theory,king1993theory}. Within the Berry-phase approach, the the P$_s$ is computed by considering the polarization difference between the ferroelectric phase and a centrosymmetric reference phase, which is the cubic BaTiO$_3$ in the present study. In the context of polaron systems, the considered reference system was a cubic BaTiO$_3$ with a localized charge carrier. 
\section{Results}

With our results, we attempt to unravel how electron and hole polarons couple to ferroelectricity in BaTiO$_3$, first by examining the role of $\delta_{FE}$ in the stability of the polarons (Section A), and then by analyzing the impact of these localized entities on the local ferroelectric order in the ferroelectric phases of the material (electron polarons: Section B; hole polarons: Section C).

\subsection{Polarons : Electrons and Hole Polarons in BaTiO\texorpdfstring{$_3$}{3}}

To acheive our goals, we inspect the formation of small electron and hole polaron in three polymorphs of BaTiO$_3$ - cubic, tetragonal, and rhombohedral. Analysis of the charge density plots confirmed the formation of small polarons and the spatial distribution of charge. As shown in Figure~\ref{Electron-polaron}, a Ti-3$d$ orbital character was observed across all three polymorphs for electron polarons, while an O-2$p$ orbital character was exhibited for hole polarons in all three polymorphs. For the electron polarons in the cubic phase, we observed an admixture of d$_{xy}$ and d$_{yz}$ orbital character, while the preferential occupation of the extra electron in the tetragonal phase was the d$_{xy}$ orbital. For the rhombohedral phase, electron polaron occupies the d$_{xz}$ orbital. For hole polarons, we observed an admixture of p$_{y}$ and p$_{z}$ orbital character for cubic phase, a p$_{y}$ orbital character for tetragonal phase, while a p$_{x}$ orbital character was observed for the rhombohedral phase of BaTiO$_3$. The density of states (DOS) plots for electron and hole polarons in different phases of BaTiO$_3$ are provided in the Figure S1 of supplementary materials (SM)\cite{supplemental_material}.
\begin{figure}[hbt]
\includegraphics[width=85mm,scale=1]{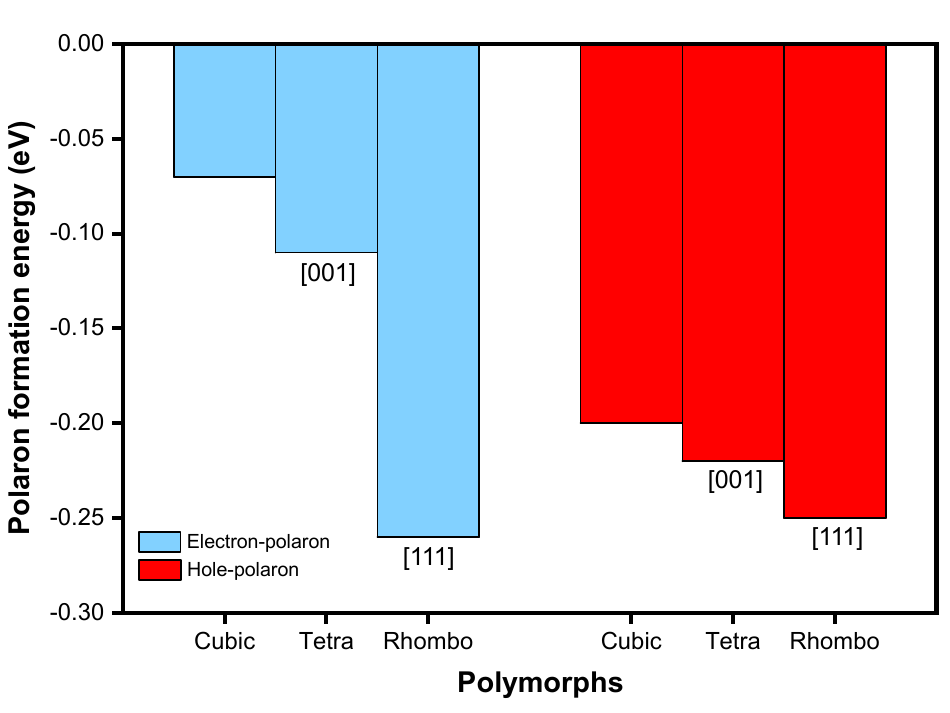}
\caption{The comparison of polaron formation energy for different phases of BaTiO$_3$ for both the electron and hole polaron. The figure depicts the enhanced stability of electron polaron for rhombohedral phase, while the same is not observed in the case of hole polaron, indicating the effect of [111] off-centering of Ti atom in stabilizing the polaron. The polarization direction is indicated within square brackets and is not shown in the case of the cubic phase because it is non-polar.}
\label{fig:epolcomparison}
\end{figure}
\vspace{3mm}

Further insights on polaron stability can be obtained from the dependence of E$_{POL}$ on the crystal symmetry. The stability of electron polarons in BaTiO$_{3}$ exhibits a clear trend with crystal symmetry, increasing from cubic to tetragonal to rhombohedral phases as shown in Figure~\ref{fig:epolcomparison}. This can be attributed to two reasons: first, the level of antibonding hybridization between the Ti-3$d$ and O-2$p$ orbitals at the conduction band minimum and second, the degree of Ti off-centering. The rhombohedral BaTiO$_{3}$ exhibits an E$_{POL}$ of -0.26 eV, which is the lowest of all the three phases considered. The cubic and tetragonal phases show an E$_{POL}$ of -0.07 eV and -0.11 eV respectively, clearly showing how rhombohedral phase outperforms the other two phases. As mentioned previously, this is attributed to the strong anti-bonding hybridization in this phase, thus allowing maximum energy release upon electron localization on the Ti-3$d$ orbital. A large electronic gain energy (E$_{EL}$) of -0.49 eV compared to the tetragonal phase's -0.34 eV validates this. The decrease in $\delta_{FE}$ of the polaronic octahedron is more significant for the rhombohedral phase than for the tetragonal phase (discussed in detail in the next section), which further validates our observation because the $\delta_{FE}$ in BaTiO$_{3}$ is strongly associated with the hybridization between Ti-3$d$ and O-2$p$. Furthermore, the FE tetragonal and rhombohedral phases, because they already possess off-centering, require less energy to distort compared to the cubic phase. This is clear from the structural energy cost (E$_{ST}$), which is calculated to be 0.27 eV from cubic, while for the tetragonal and rhombohedral phase, this is calculated to be 0.23 eV. 
These results align with previous findings by Tsunoda et al., highlighting the importance of anti-bonding hybridisation and local displacement of Ti ions along the [111] direction in both rhombohedral BaTiO$_3$ and structurally disordered cubic phases for stabilizing self-trapped polarons.~\cite{tsunoda2019stabilization}

\vspace{3mm}

Interestingly, this trend is not exactly mimicked by hole polarons, as they are already evidently stable in all three polymorphs. Polaron formation energies of -0.20 eV, -0.22 eV, and -0.25 eV are recorded for cubic, tetragonal and rhombohedral BaTiO$_3$, respectively. Although the trend of an increase in stability as one transitions from cubic to rhombohedral is also observed for the hole polaron system, the differences are modest, as shown in Figure~\ref{fig:epolcomparison}.  Unlike electron polarons, holes are localized on O-2$p$ bonding orbitals, and stabilization involves weakening the bonding O-2$p$ and Ti-3$d$ hybrids. Because the bonding states are already lower in energy and more stable, the hole-polaron formation results in a lower energy gain and smaller variations in the hole-polaron stability. Symmetry also plays a smaller role in the case of hole polarons because the distortion of the O atom, on which the hole polaron forms, from its centrosymmetric position in the ferroelectric phases is computed to be approximately 90\% less than that of Ti distortion. The combined effect could possibly explain why a small phase-dependent variation in E$_{POL}$ was observed in the case of hole polarons. The slight increase in E$_{POL}$ can be attributed to the pre-distorted TiO$_{6}$ octahedron in the ferroelectric phases, which could slightly ease the relaxation of the O-site.  
\begin{figure*}[hbt]
\includegraphics[width=0.9\textwidth]{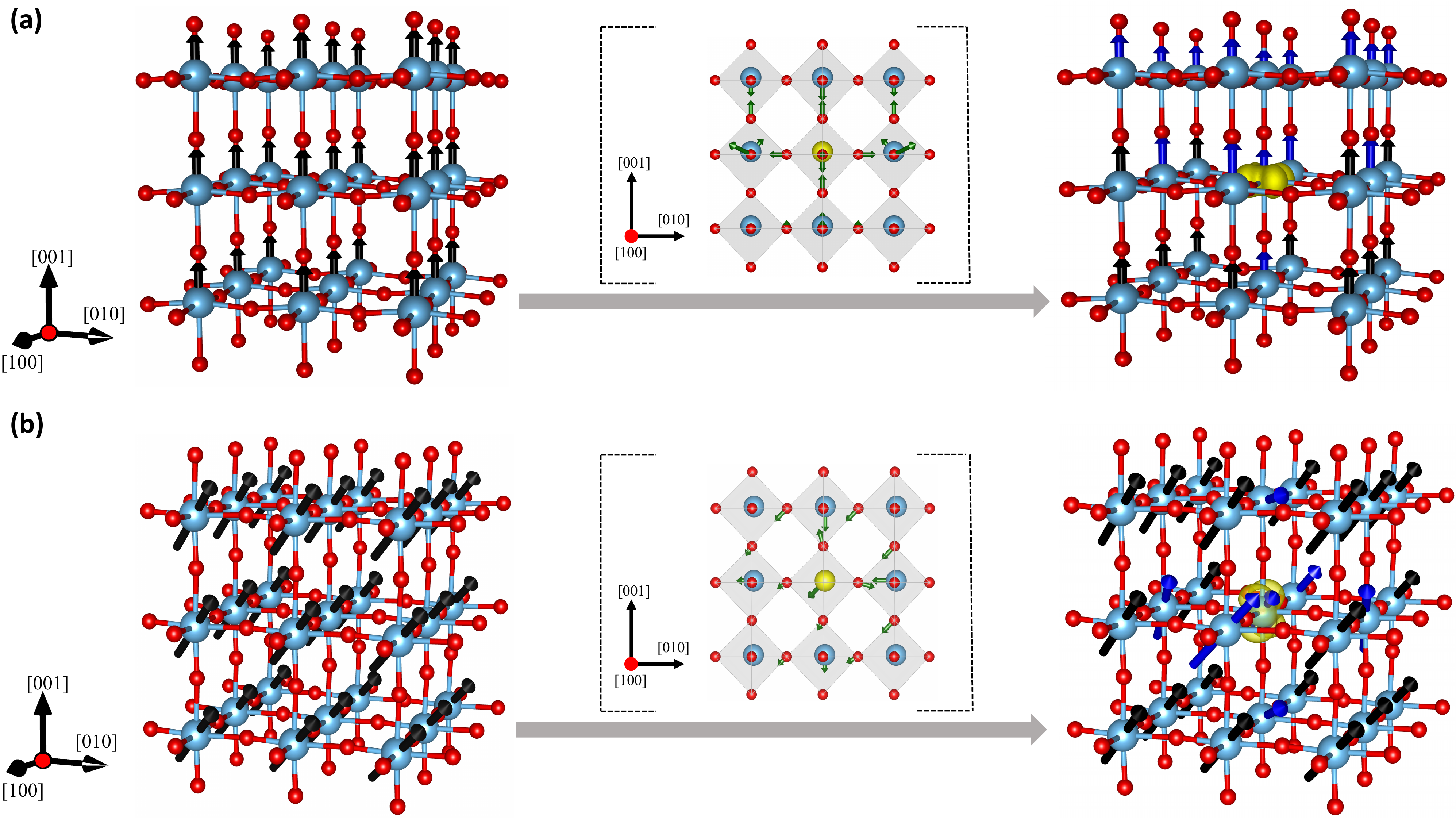}
\caption{The figure represents a schematic depiction of the evolution of $\delta_{FE}$ from a pristine to polaronic state as an electron localizes on a Ti atom in the (a) tetragonal and (b) rhombohedral phases of BaTiO$_3$. The blue arrows in the figures on the right indicate the ferroelectric distortions, which vary in both magnitude and direction compared to the $\delta_{FE}$ in the pristine state, represented by black arrows. These transformations in the $\delta_{FE}$ are attributed to the polaronic distortions, as represented in square brackets. Here, the atom represented in yellow colour is the polaronic site and the green arrows represent the polaronic distortion. Table S1 and S2 of the supplementary material provides the changes to $\delta_{FE}$ in more detail.~\cite{supplemental_material}}
\label{Ferroelectric-polaronic distortion for electron polaron}
\end{figure*}
\vspace{2mm}

Furthermore, we observed phase- and polaron-dependent modifications to the Ti-O bond lengths induced by polaron formation, providing a structural foundation for the detailed $\delta_{FE}$ analysis in Sections B-C. The formation of electron polarons generally elongates the bonds within the polaronic octahedron, consequently leading to compensatory contractions in the nearest-neighbor bonds, with distinct patterns observed for each phase. Similarly, the Ti-O bond lengths were affected by hole polaron formation, with different elongation patterns observed for different phases. The bond-length modifications for electron and hole polarons in the tetragonal and rhombohedral phases are detailed separately in Tables S1 to S4 of the SM\cite{supplemental_material}.  These modifications to the bonds form the basis for the site-specific $\delta_{FE}$ changes discussed in the following sections. 

\subsection{Effect of electron polarons on the ferroelectric properties}

Having established the influence of ferroelectricity on polaron formation, we now examine how electron and hole polarons affect the ferroelectric properties of tetragonal and rhombohedral BaTiO3. All results below are based on calculations performed with a 3x3x3 supercell.

\vspace{3mm}

In tetragonal BaTiO$_3$, the extra electron is localized in the d$_{xy}$ orbital of the Ti site, increasing the in-plane electron density and thereby elongating the in-plane Ti-O bonds owing to the enhanced electron-electron repulsion with the in-plane O atoms. Concurrently, the redistribution of the electron density leads to reduced hybridization between the polaronic Ti atom and out-of-plane O atoms, which weakens the out-of-plane Ti-O bonds. This results in cooperative atomic shifts: the polaronic Ti atom displaces $\sim$0.014~\AA\ towards the octahedral center, in-plane O atoms shift outward ($\sim$0.07~\AA), and out-of-plane O atoms move $\sim$0.02~\AA\ along the [001] direction (Figure~\ref{Ferroelectric-polaronic distortion for electron polaron}(a)). Consequently, these distortions result in a 15\% decrease in $\delta_{FE}$ in the polaronic octahedron and push the unit cell into a cubic-like symmetry. Polaron-induced distortions propagate anisotropically to the nearest neighboring sites, with the distortion pattern influenced by the intrinsic [001] polarization of the system: The 1NN above the polaronic octahedron shows a substantial decrease in $\delta_{FE}$, while those below experience only minor suppression because the tetragonal axis hinders the distortion opposite to the polarization. This distinction aligns with the atomic relaxations depicted in Figure~\ref{Ferroelectric-polaronic distortion for electron polaron}(a), where the Ti atom above the polaron shifts downward noticeably more than the Ti atom below. The 1NN octahedra in the same x-y plane showed a modest increase in $\delta_{FE}$. Table S1 of the SM details the quantitative data of the polaron-induced $\delta_{FE}$ changes\cite{supplemental_material}.

\begin{table*}[t]
\centering
\caption{\label{tab:table1}%
Macroscopic polarization values, relative energies, and magnetic moments for pristine, electron polaron and hole polaron states for different structural phases (Cubic, Tetragonal, and Rhombohedral) of BaTiO$_3$. The relative energies correspond to the energy differences between the cubic and ferroelectric phases for the pristine systems and between the cubic polaron and ferroelectric polaron systems for the polaron cases. The experimental values of P$_s$ noted for the tetragonal and rhombohedral phases of pristine BaTiO$_3$ are 0.26 C/m$^2$ and 0.34 C/m$^2$ respectively~\cite{wieder1955electrical, hewat1973structure}}
\label{tab:polaron_properties}
\begin{tabular}{llccc}
\toprule
Phase & System & Spontaneous Polarization (C/m$^2$) & Relative Energy (eV/f.u.) & Magnetic Moment ($\mu_B$) \\
\midrule
\multirow{3}{*}{Cubic} 
 & Pristine         & 0.00  & ---    & 0.00 \\
 & Electron Polaron & 0.00 & --- & 0.94 \\
 & Hole Polaron     & 0.00 & --- & 0.92 \\
\midrule
\multirow{3}{*}{Tetragonal} 
 & Pristine         & 0.39  & 0.030  & 0.00 \\
 & Electron Polaron & 0.38 & 0.032 & 0.95 \\
 & Hole Polaron     & 0.395 & 0.035 & 0.926 \\
\midrule
\multirow{3}{*}{Rhombohedral} 
 & Pristine         & 0.45  & 0.044  & 0.00 \\
 & Electron Polaron & 0.4 & 0.048 & 0.95 \\
 & Hole Polaron     & 0.46 & 0.038 & 0.912 \\
\bottomrule
\end{tabular}
\end{table*}

Upon the formation of electron polaron in the rhombohedral phase of BaTiO$_3$, pronounced anisotropic distortions are observed owing to a stronger polaron-ferroelectric coupling. The Ti site which holds a C$_{3v}$ site symmetry in the rhombohedral phase, creates antibonding states with the neighboring O-2$p$ orbitals at the conduction band minimum. As discussed in Tsunoda et al.’s study, electron localization on a Ti-3$d$ orbital modifies the TiO$_6$ octahedra such that it reduces the antibonding hybridization with neighboring O-2$p$ orbitals\cite{tsunoda2019stabilization}. This observation is similar to our calculations (in our case, polaronic Ti atom shifting towards the center of octahedron by $\sim$0.06~\AA), and given that such reduced hybridization indicates reduced ferroelectricity for BaTiO$_3$, this can explain the significant reduction of [111] $\delta_{FE}$ by 68\% observed in the polaronic octahedron. The preexisting [111] polarization imposes a strong directional anisotropy on the atomic shifts, as illustrated in Figure~\ref{Ferroelectric-polaronic distortion for electron polaron}(b). The O atoms (of the polaronic octahedron) in the direction of polarization ([100], [010], and [001] directions) shifted significantly by $\sim$0.09~\AA, while the O atoms opposite to the direction of polarization ([$\bar{1}00$], [$0\bar{1}0$], [$00\bar{1}$]) displayed a comparitively minimal distortion of $\sim$0.01~\AA. The polarization-biased electron-lattice coupling is reflected by this asymmetry, where the aligned O atoms experience stronger repulsive forces from the localized electron. Furthermore, the displaced O atoms within the polaronic octahedron exhibited an interesting direction-specific pattern, shifting predominantly along their directional axis. This directional anisotropy is also inherited by the neighboring octahedra, where significant changes in $\delta_{FE}$ occur at the nearest-neighbor sites aligned with the polarization direction, compared to smaller changes at sites opposite to it. Quantitative data supporting these trends are provided in Table S2 of the SM\cite{supplemental_material}. 

\begin{figure*}[hbt]
\includegraphics[width=0.9\textwidth]{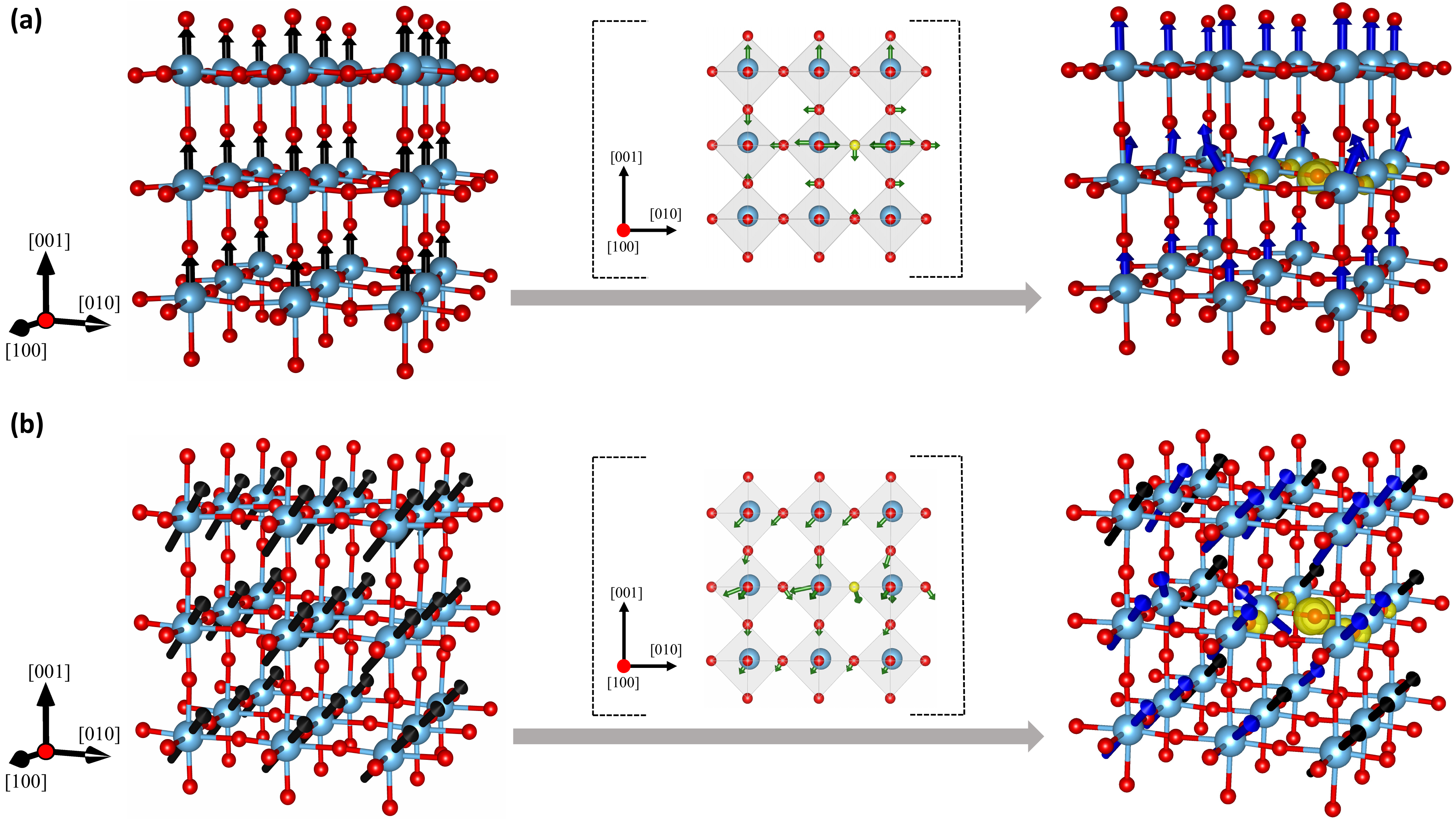} \caption{The figure represents a schematic depiction of the evolution of $\delta_{FE}$ from a pristine to polaronic state as a hole localizes on an O atom in the (a) tetragonal and (b) rhombohedral phases of BaTiO$_3$. The blue arrows in the figures on the right indicate the $\delta_{FE}$, which vary in both magnitude and direction compared to the $\delta_{FE}$ in the pristine state, represented by black arrows. These transformations in the $\delta_{FE}$ are attributed to the polaronic distortions, as represented in square brackets. Here, the yellow coloured atom represents the polaronic site and the green arrows depict the polaronic distortion. Table S3 and S4 of the supplementary material provides the changes to $\delta_{FE}$ in more detail.~\cite{supplemental_material}}
\label{Ferrolectric-polaronic distortion due to hole localization}
\end{figure*}
\vspace{3mm}

This directional bias was systematically reversed when the $\delta_{FE}$ was inverted in the initial symmetry of both phases, whereas such directional anisotropy was absent in the cubic phase, confirming the role of the inherent polarization field in mediating polaronic distortions. Owing to the limited spatial range of the polaron-induced $\delta_{FE}$ pattern, the overall P$_{s}$ remained near-pristine, whereas the ferroelectric-paraelectric energy differences exhibited minor shifts, as listed in Table 1. The self-trapping of electron polarons is also accompanied by the introduction of magnetism (0.94–0.95 $\mu{B}$) into an otherwise nonmagnetic system. The reduction of normal Ti$^{4+}$ to Ti$^{3+}$ at the polaronic site introduces an unpaired Ti-3$d$ spin-polarized electron, resulting in magnetism that is highly concentrated at the polaronic site. Such polaron-induced magnetism in inherently ferroelectric systems opens up the possibility of multiferroism, which can be investigated in future studies.

\subsection{Effect of hole polarons on the ferroelectric properties}

For both tetragonal and rhombohedral BaTiO$_3$, hole polaron formation leads to nuanced, but distinct, local ferroelectricity modifications. Upon local analysis of the hole polaron in the tetragonal phase, the distortions seem to reveal a planar dependence: in the x-y plane containing the polaronic O atom, the 1NN Ti atoms distort in opposite y-directions (by $\sim$0.08~\AA), which induces new [010] local dipoles in these octahedra (Figure~\ref{Ferrolectric-polaronic distortion due to hole localization}(a)). Because the induced dipoles are equal and opposite, they cancel each other out and do not contribute to net polarization. The surrounding O atoms in the x-y plane shift slightly (by $\sim$0.02~\AA) towards the polaronic site owing to the reduced electron-electron repulsion with the polaronic O atom. These shifts, confined to the plane, also introduce additional minor [110] distortions in their respective octahedra; however, they cancel out owing to symmetry. The changes in $\delta_{FE}$ for the neighboring sites are listed in detail in Table S3 of the SM\cite{supplemental_material}. For the neighboring Ti atoms above the polaronic plane, we observed a modest increase in off-centering in the [001] direction (by $\sim$0.01~\AA), resulting in a 6\% increase in the local $\delta_{FE}$. However, for the Ti atoms in the plane below, the distortions were insignificant (Figure~\ref{Ferrolectric-polaronic distortion due to hole localization}(a)), and the resulting changes in $\delta_{FE}$ were minimal (3\% decrease). Again, such a discrepancy could be due to the influence of the pre-existing tetragonal polarization.

\vspace{3mm}

Our calculations revealed more complex and anisotropic alterations to $\delta_{FE}$ upon the formation of hole polarons in the rhombohedral phase of BaTiO$_3$, as depicted in Figure~\ref{Ferrolectric-polaronic distortion due to hole localization}(b). The 1NN Ti atoms undergo asymmetric distortions; one exhibits a minimal shift, while  the other undergoes a substantial shift of $\sim$0.17~\AA\ along the [$0\bar{1}0$], resulting in the reversal of the local [010] polarization at that site. Furthermore, the polaronic O atom experienced displacements of $\sim$0.015 and $\sim$0.03~\AA\ along the [010] and [$\bar{1}00$] directions, respectively. These polaronic distortions resulted in significant changes in the Ti-O bonds and local $\delta_{FE}$ (detailed in Table S4 of SM) around the polaron, including a reduced 60\% negative $\delta_{FE}$ (for [010] polarization) for the 1NN octahedron along the [$0\bar{1}0$] direction of the polaron. The other most affected octahedron is the 2NN along the same direction, where we observed a $\sim$75\% reduction (for the [010] $\delta_{FE}$ ), primarily due to the significant distortion (by $\sim$0.09~\AA) of the 2NN Ti atom along the [$0\bar{1}0$] direction. As observed in the case of electron polarons, the hole polaron distortions also revealed a directional dependence governed by the intrinsic polarization direction, particularly for the 2NN TiO$_6$ octahedra in the [010] and [001] directions. The changes in these sites are clearly larger than the minimal modifications at the sites opposite to the polarization. These changes in $\delta_{FE}$ are detailed in Table S4 of the SM and are depicted schematically in Figure~\ref{Ferrolectric-polaronic distortion due to hole localization}(b)\cite{supplemental_material}. 

Consistent with the previously described case for electron polarons, the reversal of the initial $\delta_{FE}$ naturally leads to a reversed polaronic distortion pattern, further establishing the influence of the polarization on the behavior of the polaron. The cubic phase, also for hole polarons, shows isotropic distortions because it lacks inherent polarization. Across the phases, the overall P$_s$ and energy differences between the ferroelectric and paraelectric polaronic states remained close to the pristine values, as listed in Table 1. The hole polarons introduce local magnetic moments, as listed in Table 1, across all phases.

\vspace*{-5mm}
\section{Conclusion and Summary}

This study investigated the interplay between ferroelectricity and polaron formation in the cubic, tetragonal, and rhombohedral phases of BaTiO$_3$. The research involved understanding the effect of ferroelectricity on electron and hole polaron behavior in all three phases and how polarons, in turn, affect the ferroelectric properties of the systems. The findings reveal the stability of the polaronic solution for both hole and electron polarons across all thrpolaree polymorphs of BaTiO$_3$, with the stability of electron polarons exhibiting a strong phase dependence. The polaron formation energy peaks in the rhombohedral phase due to the significant [111] Ti off-centering and the ability to reduce anti-bonding hybridisation. However, this effect is less pronounced for hole polarons trapped at oxygen sites, as they already exhibit greater stability in all phases. The study also investigated the impact of polarons on ferroelectricity, analyzing electron and hole polarons separately. It was observed that the effect on ferroelectricity is substantial at polaron localization sites and their immediate neighborhoods, whereas for sites farther from the polaron, the effect is zero or negligible. This highlights the localized nature of polarons, which significantly affect local ferroelectricity without greatly impacting macroscopic polarization. This holds true for both electron and hole polarons, although the effect on ferroelectricity in each case differs. An interesting aspect of the current work is the observed polaron-ferroelectric coupling that influences how the $\delta_{FE}$ at different sites were modified. A detailed examination revealed that the polarization field plays a significant role in polaron-induced distortions, determining the extent of modification in ferroelectric distortions due to their strong coupling. In the electron polaron tetragonal system, planar anisotropic distortions are evident, with more pronounced effects on atoms aligned with the polarization direction compared to those in the opposite direction. Similarly, in the electron polaron rhombohedral system, polaronic shifts display stronger anisotropy, with atoms aligned with each polarization component experiencing greater shifts than those in the opposite direction. Hole polarons exhibit a more complex coupling but still show anisotropy in polaronic distortions influenced by the polarization field. As polaron formation is associated with a magnetic moment, ferroelectricity and magnetism coexist in these systems. By employing the HSE06 functional to address the limitations of DFT+U, we aimed to propose a framework that, with further refinement, could potentially lead to multiferroicity in the ferroelectric phases of BaTiO$_3$.

\section{ACKNOWLEDGMENTS}

This research was supported by the National Recovery and Resilience Plan (NRRP), Mission 4 Component 2 Investment 1.3 - Project NEST (Network 4 Energy Sustainable Transition) of Ministero dell’Università e della Ricerca (MUR), funded by the European Union – NextGenerationEU. The work was funded partially by the Austrian Science Fund (FWF) 10.55776/F81 project TACO. The computational results were achieved using the Vienna Scientific Cluster (VSC).

\bibliography{references}

\end{document}